\documentclass[12pt]{article}%
\usepackage[nosort]{cite}
\usepackage[usenames, dvipsnames]{xcolor}
\usepackage{graphicx}
\usepackage{multicol}
\usepackage{amsfonts}
\usepackage{amssymb}
\usepackage{amsmath}
\usepackage{heck}
\usepackage{afterpage}
\usepackage{setspace}
\usepackage{verbatim}
\usepackage{color}
\usepackage{longtable}
\usepackage{float}
\usepackage{subcaption}
\usepackage{epsfig}
\usepackage{enumerate}
\usepackage{epstopdf}
\usepackage[enableskew, vcentermath]{youngtab}
\usepackage{adjustbox}
\usepackage{multirow}
\usepackage{tikz}
\usepackage[margin=1in]{geometry}
\usepackage{titletoc}
\usepackage{hyperref}
\usepackage[percent]{overpic}
\usepackage{gensymb}
\usepackage{mathtools}
\usepackage{tikz-cd}%
\providecommand{\U}[1]{\protect\rule{.1in}{.1in}}
\pdfoutput=1
\newsavebox{\mysavebox}

\hypersetup{colorlinks,citecolor=black,filecolor=black,linkcolor=black,urlcolor=black}
\usetikzlibrary{decorations.markings}

\numberwithin{equation}{section}

\usetikzlibrary{chains}
\allowdisplaybreaks
\tikzset{node distance=2em, ch/.style={circle,draw,on chain,inner sep=2pt},chj/.style={ch,join},every path/.style={shorten >=4pt,shorten <=4pt},line width=1pt,baseline=-1ex}

\newcommand{\ba}{\begin{eqnarray}}
\newcommand{\ea}{\end{eqnarray}}

\newcommand{\be}{\begin{equation}}
\newcommand{\ee}{\end{equation}}

\tikzstyle{startstop} = [rectangle, rounded corners, minimum width=3cm, minimum height=1cm,text centered, draw=black, fill=blue!10]
\tikzstyle{startstop} = [rectangle, rounded corners, minimum width=3cm, minimum height=1cm,text centered, draw=black, fill=blue!10]
\tikzstyle{io} = [trapezium, trapezium left angle=70, trapezium right angle=110, minimum width=3cm, minimum height=1cm, text centered, draw=black, fill=blue!30]
\tikzstyle{process} = [rectangle, minimum width=3cm, minimum height=1cm, text centered, draw=black, fill=orange!30]
\tikzstyle{decision} = [diamond, minimum width=3cm, minimum height=1cm, text centered, draw=black, fill=green!30]
\tikzstyle{arrow} = [thick,->,>=stealth]
\tikzset{->-/.style={decoration={
  markings,
  mark=at position #1 with {\arrow[scale=2.4]{>}}},postaction={decorate}}}
\makeatletter \@addtoreset{equation}{section} \makeatother

\begin{document}


\date{July 2020}

\title{Qubit Construction in 6D SCFTs}

\institution{PENN}{\centerline{Department of Physics and Astronomy, University of Pennsylvania, Philadelphia, PA 19104, USA}}

\authors{
Jonathan J. Heckman\footnote{e-mail: {\tt jheckman@sas.upenn.edu}}
}

\abstract{We consider a class of 6D superconformal field theories (SCFTs)
which have a large $N$ limit and a semi-classical gravity dual description.
Using the quiver-like structure of 6D SCFTs
we study a subsector of operators protected from large operator
mixing. These operators are characterized by degrees of freedom in a one-dimensional spin chain,
and the associated states are generically highly entangled.
This provides a concrete realization of qubit-like states in a strongly coupled quantum field theory.
Renormalization group flows triggered by deformations of 6D UV fixed points
translate to specific deformations of these one-dimensional spin chains. We also present a conjectural
spin chain Hamiltonian which tracks the evolution of these states as a function of renormalization group
flow, and study qubit manipulation in this setting. Similar considerations hold for
theories without $AdS$ duals, such as 6D little string theories and 4D SCFTs obtained from compactification of the partial
tensor branch theory on a $T^2$.}

\maketitle

\setcounter{tocdepth}{2}



\newpage

\section{Introduction \label{sec:INTRO}}

There are intriguing connections between
between information theoretic structures and the emergence of
spacetime. An undergirding assumption in this setup is that entangled ``qubits''
in a conformal field theory (CFT) serve to encode the structure of a bulk
gravitational dual. There are by now a number of suggestive holographic
structures in this setting as found in references \cite{Maldacena:2001kr, Ryu:2006bv, Ryu:2006ef, Hubeny:2007xt,
VanRaamsdonk:2010pw}. There are also
proposed connections to phenomena from condensed matter and quantum information
systems such as references \cite{Swingle:2009bg, Swingle:2012wq, Nozaki:2012zj, Qi:2013caa, Hartman:2013qma,
Roberts:2014isa, Balasubramanian:2014hda, Marolf:2015vma, Pastawski:2015qua, Yang:2015uoa, Peach:2017npp,
Balasubramanian:2018por}.

This also raises some additional questions. For example, to make further tests
of these proposals it would be nice to have some explicit examples of qubits
in interacting conformal field theories with genuine stringy holographic
duals. Additionally, there is a suggestive geometric \textquotedblleft lattice
structure\textquotedblright\ built into many discussions connecting holography and
tensor networks, but this also poses some puzzles for the role of
the lattice suggested by a tensor network and its connection
to spacetime locality.

Our aim in the present note will be to construct some explicit examples
of qubit systems in superconformal field theories (SCFTs). Perhaps surprisingly,
we find it simplest to construct these states in 6D SCFTs, as well as in their dimensional reduction
on a $T^2$. The class of theories we consider resemble generalized quiver gauge theories which
take the form:%
\begin{equation}
\lbrack G_{0}]-G_{1}-...-G_{N-1}-[G_{N}].\label{genquiver}%
\end{equation}
Here, each $G_{i}$ for $i = 1,...,N$ denotes a gauge symmetry factor, with $G_0$ and $G_N$ flavor symmetry factors.
The links between these gauge groups correspond to \textquotedblleft conformal
matter,\textquotedblright\ as in references \cite{DelZotto:2014hpa, Heckman:2014qba}.

The main class of 6D\ SCFTs we consider arise from the worldvolume of $N$
M5-branes filling $\mathbb{R}^{5,1}$ and probing the transverse geometry $\mathbb{R}_{\bot}\times
\mathbb{C}^{2}/\Gamma_{ADE}$ where $\Gamma_{ADE}\subset SU(2)$ is a finite
subgroup. The subscript here serves to remind us that there is an
ADE\ classification of such finite subgroups, and these are in one to one
correspondence with the ADE\ series of Lie algebras. Indeed, M-theory on
$\mathbb{R}^{6,1} \times \mathbb{C}^{2}/\Gamma_{ADE}$ engineers 7D\ Super Yang-Mills theory with the
corresponding gauge group. Placing M5-branes on top of an ADE singularity but
separated from each other in the transverse $\mathbb{R}_{\bot}$ direction
leads to a quiver-like structure as in line (\ref{genquiver}) with gauge group
factors $G_{i}=G_{ADE}$, with the strength of each gauge coupling depending inversely
on the separation length between neighboring M5-branes. The
\textquotedblleft matter\textquotedblright\ of the generalized quiver
corresponds to degrees of freedom localized on each M5-brane. We reach a
non-trivial conformal fixed point by making all M5-branes coincide, which
corresponds to a point of strong coupling in the moduli space of the 6D\ field
theory. In the large $N$ limit, we also achieve a holographic dual description
in M-theory given by the spacetime $AdS_{7}\times S^{4}/\Gamma_{ADE}$, with
orbifold fixed points at the north and south poles of the $S^{4}$. These fixed
points correspond to the left and right flavor symmetry factors $G_{0}$ and
$G_{N}$. There are similar long quivers without a semi-classical
gravity dual \cite{Heckman:2013pva, Heckman:2015bfa}.
One can also produce 6D\ little string
theories (LSTs) by gauging a diagonal subgroup of $G_{0}\times G_{N}$, as in
reference \cite{Bhardwaj:2015oru}. This
last class of examples produces a non-$AdS$ holographic dual with a linear
dilaton background \cite{Aharony:1998ub}.

The main message of recent work in classifying 6D\ SCFTs
(see \cite{Heckman:2013pva, Heckman:2015bfa} and \cite{Heckman:2018jxk} for a review)
is that the vast majority of such theories resemble quiver
gauge theories, and in fact all known theories can be obtained through a
process of fission and fusion \cite{Heckman:2015bfa, Heckman:2016ssk, Heckman:2018pqx}
built from such quiver-like structures. So, lessons learned here apply to a broad class of 6D\ theories
and their lower-dimensional descendants obtained from further compactification.

To construct some explicit qubits, we make use of the recent work of reference
\cite{BHL} which found that for operators with large R-charge, there are nearly
protected operator subsectors which only mix with themselves. That this is
possible at strong coupling follows from the fact that at large R-charge,
there is a further suppression in operator mixing which goes inversely in the
R-charge, much as in references \cite{Berenstein:2002zw, Hellerman:2015nra}.

Using this fact, reference \cite{BHL} identified a class of gauge invariant
local operators of the
form:%
\begin{equation}
\mathcal{O}_{m_{1}...m_{N}}=X_{1}^{(m_{1})}...X_{i}^{(m_{i})}...X_{N}%
^{(m_{N})},
\end{equation}
where we view the $X_{i}^{(m_i)}$ as bifundamental operators between neighboring pairs of groups:
\begin{equation}
G_{i-1} \overset{X_{i}^{(m_i)}}{\longrightarrow} G_{i}.
\end{equation}
We note that our indexing convention here is slightly different from \cite{BHL}.
These operators are constructed on the partial tensor branch of the 6D SCFT, and to reach the conformal fixed point one must
take a further decoupling limit in which momentum transverse to the stack of M5-branes is set to zero. This imposes
a mild condition on the spectrum of quasi-particle excitations in the 1D lattice of spins, and for the
most part we will keep this point implicit in what follows. See reference \cite{BHL} for details.

Treating the $G_{i-1}$ and $G_{i}$ as flavor symmetries, the $X_{i}$ define ``conformal matter'' operators
which have non-trivial scaling dimension
$\Delta_{X}$ and transform in a spin $s_{X}$
representation of the $SU(2)_{\mathcal{R}}$ R-symmetry.\ In the special case
where all the $G_{i}=SU(K)$, $s_{X}=1/2$ but for the D- and E-type theories,
the spin is higher \cite{Heckman:2014qba, BHL}:%
\begin{equation}%
\begin{tabular}
[c]{|c|c|c|c|c|c|}\hline
& $A_{K}$ & $D_{K}$ & $E_{6}$ & $E_{7}$ & $E_{8}$\\\hline
$s_{X}$ & $1/2$ & $1$ & $3/2$ & $2$ & $3$\\\hline
\end{tabular}
.
\end{equation}
The key feature found in \cite{BHL} is that for the composite gauge invariant operators
$\mathcal{O}_{m_{1}...m_{N}}$, the action
of the one-loop Dilatation operator on the $\mathcal{O}_{m_{1}...m_{N}}$ is simply
that of a 1D spin chain Hamiltonian. In the case where we
have A-type gauge groups with $G_{i} = SU(K)$, the one-loop Dilatation operator takes the form:
\begin{equation}
\Delta = E^{(0)} - \lambda_{A}\underset{i=1}{\overset{N}{\sum}%
} 2 \overrightarrow{S}_{i}\cdot\overrightarrow{S}_{i+1} . \label{DELTAA}%
\end{equation}
Here, the $\overrightarrow{S}_{i}$ denote the usual angular momentum operators
in the spin $1/2$ representation and we have set $\overrightarrow{S}_{N+1} = 0$.
For a spin chain with periodic
boundary conditions we would instead set $\overrightarrow{S}_{N+1}%
=\overrightarrow{S}_{1}$. The constants $E^{(0)}$ and $\lambda_{A}>0$ were
computed in \cite{BHL}. We will not need their explicit values in what follows. The
expression (\ref{DELTAA}) is the Hamiltonian for the celebrated ferromagnetic
$XXX_{s=1/2}$ Heisenberg spin chain \cite{Heisenberg:1928mqa}. It also shows up prominently in the
context of $\mathcal{N}=4$ Super Yang-Mills theory (see e.g. \cite{Berenstein:2002jq, Minahan:2002ve, Beisert:2003yb})
which served as a motivation for reference \cite{BHL}.

In the case of the D- and E-type spin chains, similar considerations hold, but we
instead get a Hamiltonian constructed from a polynomial in the
$\overrightarrow{S}_{i}\cdot\overrightarrow{S}_{i+1}$. The precise form of
this polynomial can be fully fixed by assuming that the integrable structure
present in the A-type theories persists in this broader setting. The spectrum
of excitations can now be studied using methods such as
the algebraic Bethe ansatz (see \cite{Bethe:1931, Babujian:1983ae} and \cite{Faddeev:1996iy} for a
review). One can in principle also extend this to more general excitations of
the full superconformal algebra $\mathfrak{osp}(8^{\ast}|1)$. Another
generalization has to do with taking more general closed loops and
\textquotedblleft operator impurity insertions.\textquotedblright\ All of
these cases produce similarly rich spin chain systems which are also amenable
to the same sort of analysis.

From this starting point, we can now see the emergence of a natural system of
qubits. Our plan in this note will be to use this to build up a system of
protected qubits in a higher-dimensional SCFT. Additionally, because of the
explicit form of these qubits, we can perform computations of entanglement
entropy for these states. We can also use this same setup to track the
effects of entanglement of states under 6D renormalization group flow as well as to
construct a protocol for qubit manipulation.
The picture we arrive at is reminiscent of notions appearing in the holographic tensor
network literature though we shall not attempt to make any exact
correspondences with the statements found there.

To avoid technical complications, we mainly work with the special case of
M5-branes probing an A-type singularity. Similar considerations hold for all
the D- and E-type singularities, but with the mild caveat that we are then dealing with
higher spin representations. In what follows, we will also not dwell on the
distinction between spin chains with open boundary conditions and those with
periodic boundary conditions since we will work in a thermodynamic limit where $N$ is quite large.
The case of periodic boundary conditions occurs anyway in the study of
6D\ LSTs \cite{BHL}.

\section{Qubits in 6D\ SCFTs}

We now proceed to build a system of qubits in a 6D\ SCFT. As already
mentioned, we are interested in the class of operators $\mathcal{O}%
_{m_{1}...m_{N}}$ where $m_{i}=\pm1/2$. Working with the radially quantized SCFT,
we see that each local operator specifies a state in the Hilbert
space:
\begin{equation}
\mathcal{O}_{m_{1}...m_{N}}(0)\left\vert \text{GND}\right\rangle =\left\vert
\mathcal{O}_{m_{1}...m_{N}}\right\rangle \in\mathcal{H}_{6D}\text{.}%
\end{equation}
On the other hand, the $m_{i}$ also specify a state in a 1D\ spin chain Hilbert space:
\begin{equation}
\left\vert m_{1}...m_{N}\right\rangle \in\mathcal{H}_{1D}\text{.}%
\end{equation}
From all that we have said, $\mathcal{H}_{1D}$ defines a protected subsector
of states in the 6D\ SCFT. This is the qubit system we wish to study.

The spatial direction of the spin chain is clearly related to the $\mathbb{R}_{\bot}$
direction of the M5-brane probe theory. Note that in the holographic dual, the
spin chain direction corresponds to a great arc passing from the north pole to
the south pole of $S^{4}/\Gamma_{ADE}$.

Returning to our spin chain Hamiltonian:%
\begin{equation}
\Delta = E^{(0)} - \lambda_{A} \underset{i=1}{\overset{N}{\sum}%
} 2 \overrightarrow{S}_{i}\cdot\overrightarrow{S}_{i+1},
\end{equation}
we observe that the lowest energy states actually have a large degeneracy. To
see this, observe that the total angular momentum operator:%
\begin{equation}
\overrightarrow{S} = \underset{i=1}{\overset{N}{\sum}%
}\overrightarrow{S}_{i}%
\end{equation}
commutes with $\Delta$, namely $\left[  \overrightarrow{S},\Delta \right]  =0$.
So, we can organize our energy eigenstates into
representations of $\overrightarrow{S}$. The system is also
gapless in the sense that it costs very little energy to produce an excitation
above the lowest energy states. Note also that there is a non-relativistic dispersion
relation with energy $\epsilon(p) \sim p^{2} $, so we get a scale invariant but
non-Lorentz invariant system. Some examples of entanglement entropy calculations
for Heisenberg spin chains and deformations thereof have been carried out
in references \cite{Calabrese:2004eu, Korepin:2004zz, Popkov_2005}.

Let us now discuss in more detail the lowest energy states of the system. To begin,
take the state with all $m_{i}=1/2$. This is the highest weight state of a
spin $ N/2$ representation. Introducing $S^{\pm}=S^{x}\pm iS^{y}$, we reach
the other states with the same scaling dimension by successive applications of
the lowering operator $S^{-}$. The resulting form of the states
obtained from $M$ such spin flips are discussed in \cite{Popkov_2005} and are
given by:%
\begin{equation}
\left\vert N,M\right\rangle =\frac{1}{\sqrt{C_{M,N}}}\underset{\sigma}{\sum
}\left\vert m_{\sigma(1)}...m_{\sigma(N)}\right\rangle ,
\end{equation}
where we sum over all permutations of $M$ down spins on $N$ sites.\ Here, we
have also introduced the combinatorial factor:%
\begin{equation}
C_{M,N}=\frac{N!}{M!(N-M)!}.
\end{equation}

We claim that the resulting qubits of this ground state are highly entangled states,
and similar considerations hold for quasi-particle excitations of the spin chain.
Indeed, introducing the pure state:
\begin{equation}
\rho_{M,N}=\left\vert N,M\right\rangle \left\langle N,M\right\vert ,
\end{equation}
we get a mixed state by performing a partial trace of $N-n$ spins, not
necessarily in a single contiguous block. We then get the reduced density
matrix:\footnote{As mentioned before, we are ignoring the zero momentum constraint on quasi-particle excitations
of the spin chain. We expect this to be a subleading effect so we neglect it in the discussion which follows.}
\begin{equation}
\rho_{M,N}^{(n)}=\text{Tr}_{(N-n)}\rho_{M,N}.
\end{equation}
The entanglement entropy for this was computed in \cite{Popkov_2005} for
periodic boundary conditions. In the thermodynamic limit where $N\rightarrow
\infty$ with $M/N=p$ held fixed and $n/N\leq1/2$ also fixed, this takes the
form:%
\begin{equation} \label{eeformula}
\mathcal{S}^{(n)}=-\text{Tr}\rho_{M,N}^{(n)}\log\rho_{M,N}^{(n)}\approx\frac{1}{2}\log
n+\frac{1}{2}\log(2\pi epq),
\end{equation}
with $p+q=1$.

What is the interpretation of this in the original M-theory picture? We can
consider starting with our stack of $N$ M5-branes, and can perform a partial
trace over all but $n$ of them. Doing so, we get a highly entangled stated.
Indeed, the proof of this is that in our 1D\ system we have an entanglement
entropy proportional to $\log n$.

The pure states $\rho_{M,N}$ realize explicit examples of $N$ qubit W-states
as well as generalizations thereof, as opposed to GHZ states. For
example, in a system with $N$ qubits, we can introduce the pure states:\footnote{The original terminology applies to the three qubit case
\cite{Dur_2000}, but it clearly extends, with the caveat that there are many ways to entangle four or more qubits
\cite{Verstraete_2002}.}
\begin{align}
\left\vert \text{GHZ}\right\rangle  &  =\frac{1}{\sqrt{2}}\left(  \left\vert
\uparrow...\uparrow \right\rangle +\left\vert \downarrow... \downarrow \right\rangle \right) \\
\left\vert \text{W}\right\rangle  &  =\frac{1}{\sqrt{N}}\left(  \left\vert
\downarrow \uparrow ... \uparrow\right\rangle +...+\left\vert
\uparrow \uparrow \uparrow ... \downarrow\right\rangle \right) \label{Wstate} ,
\end{align}
and then form the corresponding density matrices. In both cases, performing a
partial trace over a single qubit leads to a non-reduced density matrix, but
we observe that after doing this, the resulting density matrix for the
GHZ\ state is separable (no further entanglement between qubits after performing
additional partial traces) whereas in
the case of the W-state entanglement persists after performing a further partial trace.
It is in this sense that the qubits we have constructed are \textquotedblleft highly entangled\textquotedblright.

\section{Entanglement and 6D\ RG\ Flows}

Starting from a conformal field theory, we can consider deformations which
trigger a flow to another conformal fixed point in the infrared (IR). In the case
of 6D\ SCFTs, the available options for supersymmetry preserving
renormalization group flows are quite limited. These are always specified by
background operator vacuum expectation values (vevs)
and are referred to as tensor branch flows and Higgs
branch flows \cite{Louis:2015mka, Cordova:2016xhm} (see also \cite{Heckman:2015ola}).
We consider both sorts of flows.

\subsection{Tensor Branch Flows}

Consider first tensor branch flows. Some examples of tensor branch flows arise
from just separating the M5-branes from one another. In the case of M5-branes
probing an A-type singularity, this turns out to be the only possibility. For
more general 6D\ SCFTs other tensor branch deformations are possible and they
are all classified by suitable K\"{a}hler deformations of the associated
F-theory model \cite{Heckman:2013pva, Heckman:2015bfa}.

\begin{figure}[t!]
\begin{center}
\includegraphics[scale = 0.5, trim = {0cm 1.0cm 0cm 2.0cm}]{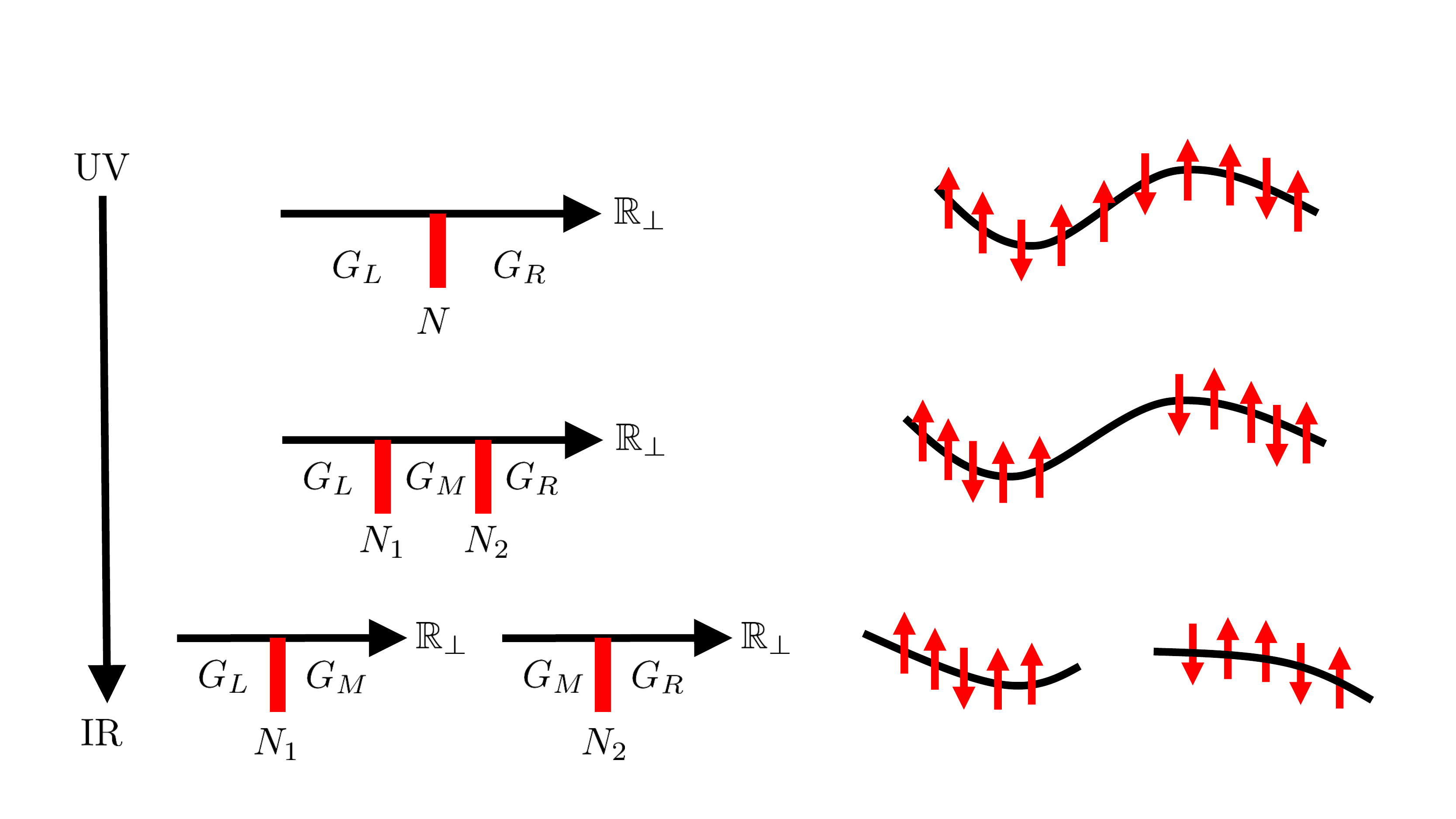}
\caption{Depiction of a tensor branch flow from the UV to the IR. In the M5-brane
picture (left) this involves separating a stack of $N = N_1 + N_2$ M5-branes into two stacks.
In the associated spin chain (right), the resulting spins separate into two decoupled sectors.
There can still be significant entanglement between the two sectors.}
\label{fig:UVIRflow}
\end{center}
\end{figure}

Sticking to the simplest case where we separate our M5-branes into two stacks
$N_{1}$ and $N_{2}$ such that $N_{1}+N_{2}=N$, we can clearly see that in the
deep infrared, our single spin chain has broken up into two independent spin
chains:%
\begin{equation}
\lbrack G_{0}]-G_{1}-...-G_{N_{1} - 1}-[G_{N_{1}}] \oplus \lbrack G_{N_{1}}]-G_{N_{1}+1}-...-G_{N - 1}-[G_{N}]
\end{equation}
where we have indicated how the gauge group $G_{N_{1}}$ has become a flavor
symmetry. In the deep infrared, this flavor symmetry acts independently on the two
decoupled SCFTs (see figure \ref{fig:UVIRflow}).

Now, we can ask about the structure of the Hilbert space of the 6D\ SCFT in
this limit. Clearly, we expect a split into two ``decoupled'' SCFTs in the infrared, so
we can write:%
\begin{equation}
\mathcal{H}_{\text{IR}}=\mathcal{H}_{N_{1}}\otimes\mathcal{H}_{\text{mix}%
}\otimes\mathcal{H}_{N_{2}}, \label{HIR}%
\end{equation}
where here, $\mathcal{H}_{\text{mix}}$ denotes the Hilbert space
of a TQFT coupled to some free fields (see e.g. \cite{Kapustin:2014gua}).

In the spin chain, we can visualize this process by working with a slightly
more general Dilatation operator / spin chain Hamiltonian:%
\begin{equation}
\Delta_{\text{IR}}= E^{(0)}_{\text{IR}} - \underset{i=1}{\overset{N}{\sum}%
}2 \lambda_{i}\overrightarrow{S}_{i}\cdot\overrightarrow{S}_{i+1},
\end{equation}
where the couplings can now be position dependent.
The limit we are discussing amounts to setting:%
\begin{align}
\lambda_{1}  &  =\lambda_{2}=...=\lambda_{N_{1} - 1},\\
\lambda_{N_{1}}  &  =0,\\
\lambda_{N_{1}+1}  &  =\lambda_{N_{1}+2}=...=\lambda_{N - 1}.
\end{align}
We note that the computation of the spin chain couplings performed in reference \cite{BHL}
displays a non-trivial dependence on $N$, so in particular when $N_{1}\neq N_{2}$, we do not
expect the couplings on the left and righthand sides of the decoupled spin
chains to be the same in the deep IR.

We can also see that there is a great deal of entanglement between the two separated
M5-brane sectors of line (\ref{HIR}). Indeed, from our discussion in the previous
section, we know that even in our spin chain subsector this scales as $\log
N_{1}$ (in the case where $N_{1}<N_{2}$). This provides evidence for the
existence of a non-trivial TQFT\ which couples these two sectors, in accord
with the general considerations presented in \cite{Heckman:2017uxe} where 6D\ SCFTs
were visualized as \textquotedblleft edge modes\textquotedblright\ of a bulk 7D\ theory
(see also \cite{Witten:1998wy, DelZotto:2015isa, Monnier:2017klz}.

We can generalize this to multiple boundaries by considering other partitions
of $N$:
\begin{equation}
N=N_{1}+...+N_{k}.
\end{equation}
In this way, we can build multi-party entangled qubits. See figure \ref{fig:MultiThroats} for a depiction.
It would be interesting to see whether this provides a higher-dimensional analog of the situation considered in
references \cite{Balasubramanian:2014hda, Marolf:2015vma, Peach:2017npp, Balasubramanian:2018por}.

\subsection{Higgs Branch Flows}

Consider next the case of Higgs branch flows. The distinguishing feature here
is that the $SU(2)_{\mathcal{R}}$ R-symmetry is broken along the flow, but a new R-symmetry
emerges in the deep infrared. The resulting class of theories which can
be achieved in these cases again resemble quiver gauge theories, but in which
there can now be different ranks of gauge groups in the generalized quiver, as
well as possible decorations by conformal matter on the left and righthand
sides \cite{Heckman:2015bfa, Heckman:2016ssk, Heckman:2018pqx}.
Importantly, in the vast majority of Higgs branch flows, the
actual number of gauge group factors again remains of order
$N$. This means that even in the associated spin chain generated in the deep IR, we again have the
same number of spins, but can now have more general position dependence:%
\begin{equation}
\Delta_{\text{IR}} = E^{(0)}_{\text{IR}}- \underset{i=1}{\overset{N }{\sum}%
} 2 \lambda_{i}\overrightarrow{S}_{i}\cdot\overrightarrow{S}_{i+1}.
\end{equation}

Indeed, in such situations, we can ask about the structure of the \textquotedblleft
ground states\textquotedblright\ found for the ultraviolet (UV) Hamiltonian. Note, however,
that $\overrightarrow{S} = \overrightarrow{S}_{1} + ... + \overrightarrow{S}_{N}$
still commutes with $H_{\text{IR}}$. So, we again have a
large degeneracy in the ground state of the spin chain.
Moreover, even though the spectrum of excitations has moved around,
the impact on the structure of the spin chain
Hilbert space is relatively mild.

\section{Interpolation}

At this point, it is interesting to ask about how to interpret the structure
of the spin chain and its Hamiltonian as we proceed from a perturbation of the
ultraviolet fixed point to a new one in the deep infrared. Here we discuss some
speculative comments in this direction.

\begin{figure}[t!]
\begin{center}
\includegraphics[scale = 0.5, trim = {0cm 1.0cm 0cm 2.0cm}]{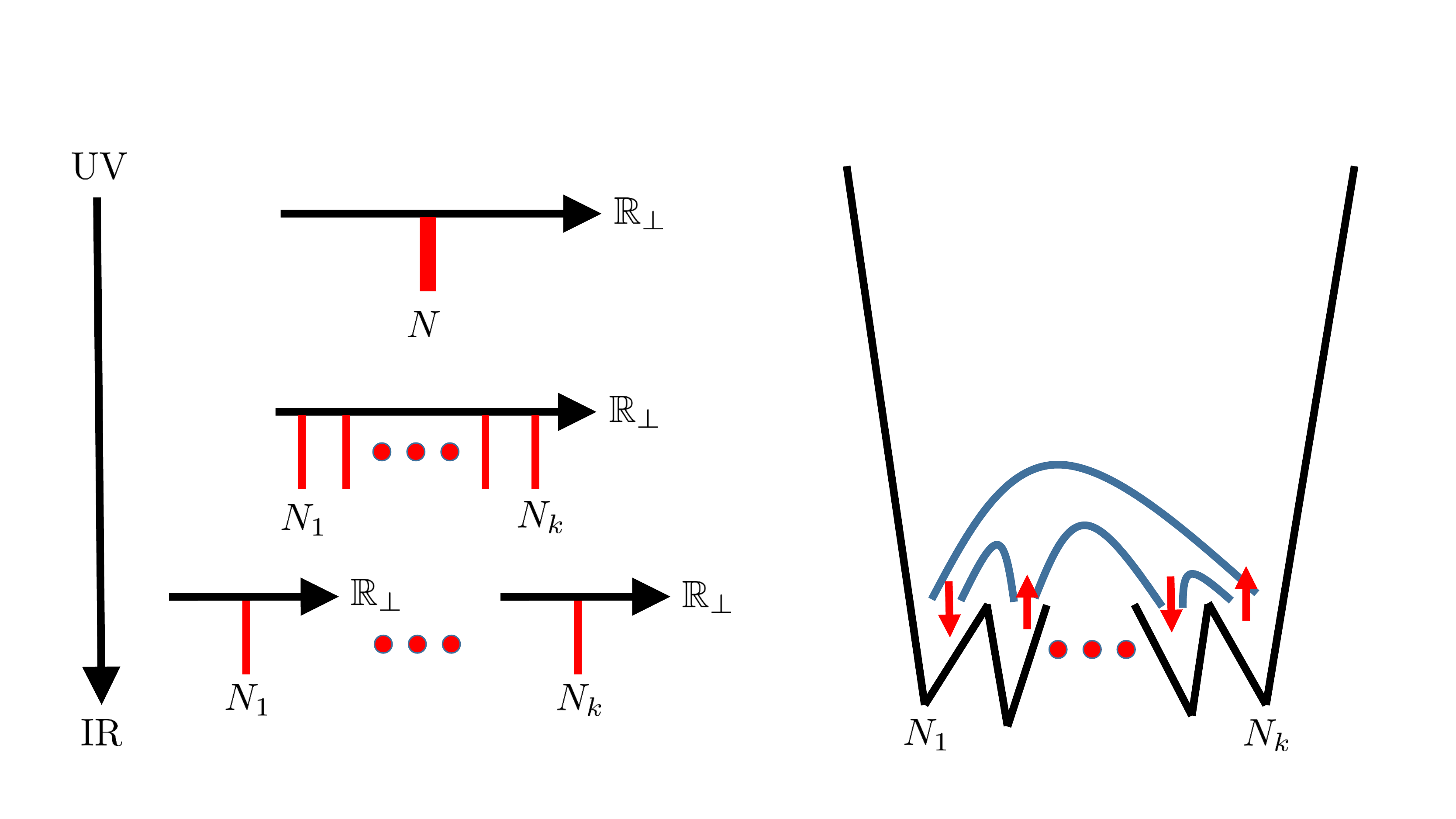}
\caption{Depiction of a multi-throat spacetime generated by pulling $N = N_1 + ... + N_k$ M5-branes
apart into separate stacks. Starting from a configuration of spins in the parent theory, we
get a multi-party entangled state in the IR theory.}
\label{fig:MultiThroats}
\end{center}
\end{figure}

First of all, we note that in the 6D\ SCFT, the spin chain Hamiltonian has the
interpretation as the one-loop Dilatation operator. Once we break conformal symmetry,
our interpretation must also be suitably loosened. That being said, it is also
clear that we can still take a state and ask how it evolves as a function of
scale. In the holographic dual setup, this corresponds to motion from the
\textquotedblleft UV\ brane\textquotedblright\ to the \textquotedblleft
IR\ brane\textquotedblright. Observe that at least for tensor branch flows,
this is immediately realized in terms of conventional branes:\ We simply take
some number of M5-branes and pass them down the throat of the AdS geometry
(see figure \ref{fig:MultiThroats}).

As a first generalization, then, we can consider a family of spin chain
Hamiltonians, one for each step in the RG\ direction. Labelling this family as
$\Delta(z)$ such that $z_{\text{UV}}$ corresponds to the UV and $z_{\text{IR}}$
corresponds to motion into the IR, we now allow our nearest neighbor interactions to depend on
RG\ time, writing $\lambda_{i}(z)$. In this context, it is appropriate to
also permit our spin operators to be $z$ dependent as well, so we write
$\overrightarrow{S}_{i}(z)$ to reflect this fact. At a given RG\ time slice,
we now can write:%
\begin{equation}
\Delta(z)=E^{(0)}(z)- \underset{i=1}{\overset{N }{\sum}}2 \lambda
_{i}(z)\overrightarrow{S}_{i}(z)\cdot\overrightarrow{S}_{i+1}(z)+....
\end{equation}
We can thus consider a slightly broader class of \textquotedblleft time
dependent\textquotedblright\ spin chains in which we evolve from
$\Delta(z_{\text{UV}})$ to $\Delta(z_{\text{IR}})$.

There are good reasons to generalize this slightly further. For one
thing, we note that as written, this Hamiltonian still preserves
$SU(2)_{\mathcal{R}}$ R-symmetry. On the other hand, we also know that at
least in Higgs branch flows, we expect the R-symmetry to be broken, only to
reemerge deep in the IR. As a further generalization, we therefore allow various
sorts of R-symmetry breaking as generated by vevs of operators in the parent UV\ theory.

In the 1D spin chain, this suggests a further generalization where we allow
$SU(2)_{\mathcal{R}}$ R-symmetry breaking terms. One option is to just
include some background magnetic field terms. We can also allow various
$SU(2)_{\mathcal{R}}$ braking terms of the sort appearing in
integrable $XYZ$ models. Including both sorts of terms, we get, in the obvious
notation:
\begin{equation}
\Delta(z)=E(z)- \underset{i=1}{\overset{N}{\sum}}%
\underset{a=1}{\overset{3}{\sum}}\left( 2 \lambda_{i}^{(a)}(z)S_{i}%
^{(a)}(z)\cdot S_{i+1}^{(a)}(z)+h_{i}^{(a)}(z)\cdot S_{i}^{(a)}(z)\right) .
\end{equation}
The main condition we need to impose is that once we
reach the IR where we recover a 6D\ SCFT, all $SU(2)_{\mathcal{R}}$ breaking
terms go to zero as $z\rightarrow z_{\text{IR}}$.

\section{Qubit Manipulation}

Given the suggestive form of our spin chain system, it is of course
interesting to ask whether we can directly manipulate the associated qubits,
and using this, probe additional structure in these systems.\footnote{We thank
A. Kar for some comments which prompted us to consider this possibility.} From
the perspective of the 6D\ SCFT, the natural operations on states include
acting with the symmetry generators of the conformal field theory, including
the Dilatation and R-symmetry operators. In particular, the Dilatation
operator corresponds to the Hamiltonian of the spin chain, governing time
evolution as a function of renormalization group scale in the 6D SCFT.
At first pass, the use of the R-symmetry generators provides a way for us to rotate qubits, but in
the interacting SCFT, we really have only the operator:
\begin{equation}
\overrightarrow{S} = \underset{i=1}{\overset{N}{\sum}%
}\overrightarrow{S}_{i},
\end{equation}
which would simultaneously manipulate many qubits all at once.

\begin{figure}[t!]
\begin{center}
\includegraphics[scale = 0.5, trim = {0cm 3.0cm 0cm 2.0cm}]{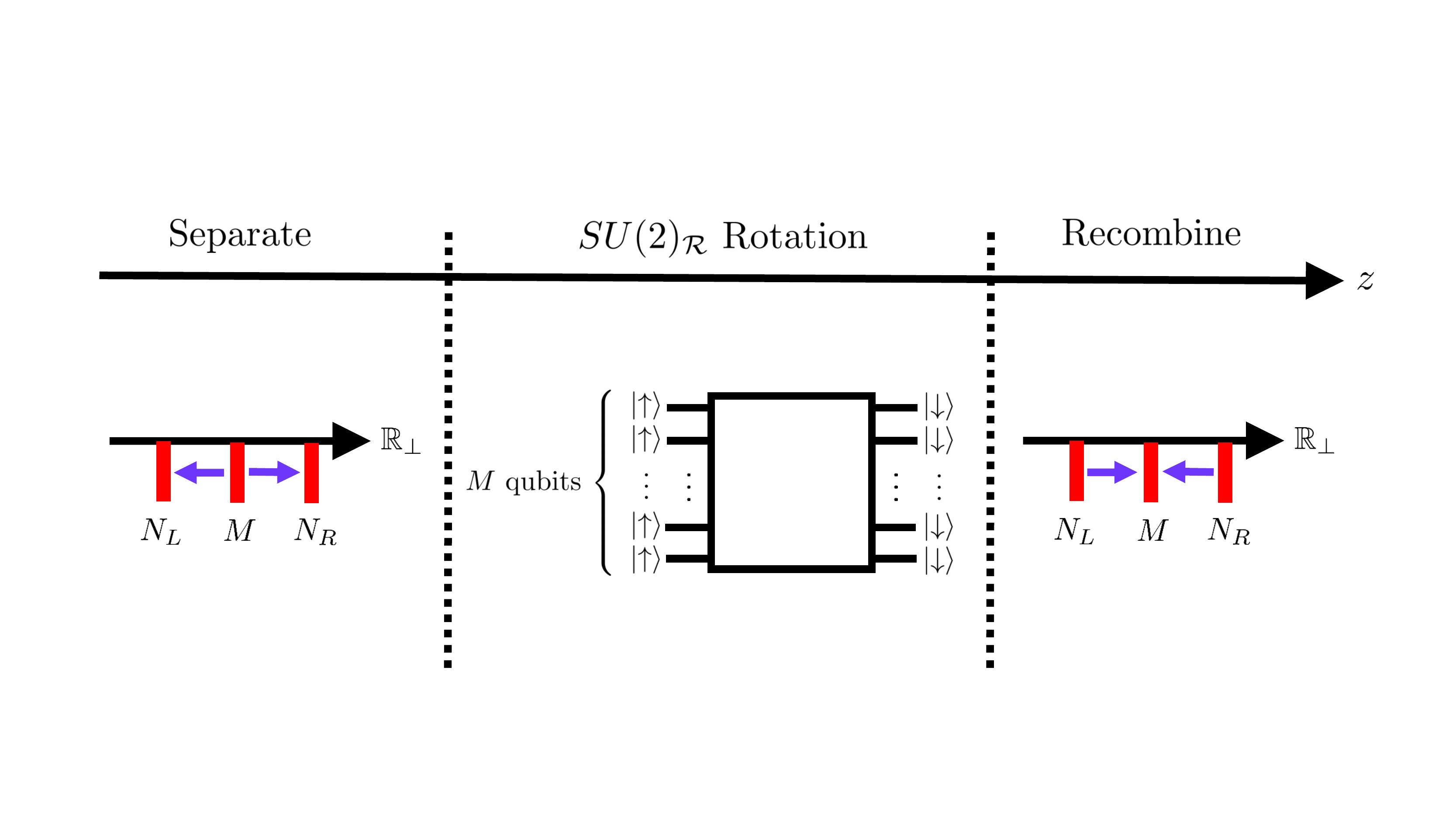}
\caption{Example of qubit manipulation as a function of RG time / trajectory time in a moduli space flow with local coordinate
$z$. The starting point is to separate M5-branes from one another. This is followed by a general Bloch sphere / $SU(2)_{\mathcal{R}}$
rotation. After this, the M5-branes are recombined. In the case of 6D SCFTs this is followed by a
projection onto the zero momentum sector of the 1D spin chain Hilbert space.}
\label{fig:QubitComputer}
\end{center}
\end{figure}

Using the M5-brane picture, however, we can see how to build a general
protocol for qubit manipulation. As a warmup, consider the 4D\ SCFT\ obtained
by compactifying the tensor branch theory on a $T^{2}$. This system has
basically the same qubit structure as the 6D\ theory, but there is no zero
momentum constraint \cite{BHL}. To manipulate an individual qubit, we consider
a deformation which involves separating a single M5-brane from all of its
neighbors in the quiver. In this limit, a link between $G_{i-1}$ and $G_{i}$
is isolated from the rest of the system, and it has
its own emergent R-symmetry in the infrared. So, we can apply a unitary $SU(2)_{\mathcal{R}}$
transformation along with an overall phase factor $\exp(i\gamma)$
$\exp(-i\overrightarrow{\theta}\cdot\overrightarrow{S})$. Doing so, we can
generate the standard single qubit operations, visualized as rotations on the
Bloch sphere. Some simple examples include the bit flip operation, or
Pauli $X$-gate:%
\begin{equation}
\sigma_{X}=\left[
\begin{array}
[c]{cc}%
0 & 1\\
1 & 0
\end{array}
\right]  =\exp(i\pi/2)\exp(-i\pi S^{(x)})\text{.}%
\end{equation}
Another example is the Hadamard gate:%
\begin{equation}
H=\frac{1}{\sqrt{2}}\left[
\begin{array}
[c]{cc}%
1 & 1\\
1 & -1
\end{array}
\right]  =\exp(i\pi/2)\exp\left(  -i\frac{\pi}{\sqrt{2}}\left(  S^{(x)}%
+S^{(z)}\right)  \right)  \text{.}%
\end{equation}

Now we can see a general way to start manipulating individual qubits: We begin
by pulling all the M5-branes away from each other. In the spin chain
Hamiltonian this corresponds to specific $z$-dependent behavior for the
nearest neighbor interactions. After they are well separated, they each have
their own emergent $SU(2)_{\mathcal{R}}$ R-symmetry and we can manipulate
their qubits individually. After this, we can bring the M5-branes back
together, and evolve further with the Dilatation operator. Note that this is a
flow in moduli space with the local coordinate of the flow playing
the role of time evolution in the qubit system. Indeed, the process of bringing the branes back together is
something which is done ``by hand'' in the brane system, though we emphasize that because this is a trajectory in
the moduli space of vacua, there is no energy cost associated with this operation.

Similar considerations clearly hold for the 6D\ SCFT system, with the mild
caveat that we need to impose the zero momentum constraint on possible
excitations. This means, for example, that after manipulating individual qubits
and bringing the stack of M5-branes back together that we need to perform a further
projection onto the zero momentum sector of the 1D spin chain Hilbert space. So,
in an actual quantum computation we would perform this operation at the very end.

As a final generalization, we can now see how to build far more involved qubit
operations. In this case we consider separating a stack of $M$ M5-branes from
the rest of the system via a tensor branch flow, manipulate that individual
set of qubits, and then bring it back to the rest of the configuration.
An example of this sort of qubit manipulation is depicted in figure
\ref{fig:QubitComputer}. It would be interesting to study further
the class of qubit operations which can be engineered in this way.

\section*{Acknowledgements}

We thank F. Baume and C. Lawrie for helpful discussions and an inspiring collaboration on
related work. We also thank M. Dierigl for helpful discussions. We thank A. Kar and O. Parrikar
for several insightful comments on an earlier draft. The work of JJH is supported
by a University Research Foundation grant at the University
of Pennsylvania.




\bibliographystyle{utphys}
\bibliography{6DQubits}

\end{document}